\documentclass[11pt,a4paper,useAMS,usenatbib]{emulateapj}
\usepackage{graphicx}

\usepackage{lineno}

\usepackage{epsfig}
\usepackage{amsmath}
\usepackage{natbib}
\usepackage{pifont}

\begin{document}

\title{Detection of a superluminous spiral galaxy \\ in the heart of a massive galaxy cluster}

\author{\'Akos Bogd\'an\altaffilmark{1}, Lorenzo Lovisari\altaffilmark{2,1}, Patrick Ogle\altaffilmark{3}, Orsolya E. Kov\'acs\altaffilmark{4,1}, \\ Thomas Jarrett\altaffilmark{5}, Christine Jones\altaffilmark{1}, William R. Forman\altaffilmark{1},  and Lauranne Lanz\altaffilmark{6}}
\affil{\altaffilmark{1}Center for Astrophysics \ding{120} Harvard \& Smithsonian, 60 Garden Street, Cambridge, MA 02138, USA; abogdan@cfa.harvard.edu}
\affil{\altaffilmark{2}INAF - Osservatorio di Astrofisica e Scienza dello Spazio di Bologna, via Piero Gobetti 93/3, I-40129 Bologna, Italy}
\affil{\altaffilmark{3}Space Telescope Science Institute, 3700 San Martin Drive, Baltimore, MD 21218, USA}
\affil{\altaffilmark{4} Department of Theoretical Physics and Astrophysics, Faculty of Science, Masaryk University, Kotl\'a\v{r}sk\'a 2, Brno, 611 37, Czech Republic}
\affil{\altaffilmark{5}Astronomy Department, University of Cape Town, Private Bag X3, Rondebosch 7701, South Africa}
\affil{\altaffilmark{6}Department of Physics, The College of New Jersey, 2000 Pennington Road, Ewing, NJ 08628, USA}
\shorttitle{SPIRAL GALAXY AT THE CENTER OF A CLUSTER}
\shortauthors{BOGD\'AN ET AL.}

\begin{abstract}
It is well established that brightest cluster galaxies (BCGs), residing in the center of galaxy clusters, are typically massive and quenched galaxies with cD or elliptical morphology. An optical survey suggested that an exotic galaxy population, superluminous spiral and lenticular galaxies could be the BCGs of some galaxy clusters. Because the cluster membership and the centroid of a cluster cannot be accurately determined based solely on optical data, we followed-up a sample of superluminous disk galaxies and their environment using  \textit{XMM-Newton} X-ray observations. Specifically, we explored seven superluminous spiral and lenticular galaxies that are candidate BCGs. We detected massive galaxy clusters around five superluminous disk galaxies and established that one superluminous spiral, 2MASX\,J16273931+3002239, is the central BCG of a galaxy cluster. The temperature and total mass of the cluster are $kT_{\rm 500}=3.55^{+0.18}_{-0.20}$~keV and $M_{\rm 500} = (2.39 \pm 0.19) \times 10^{14} \ \rm{M_{\odot}} $. We identified the central galaxies of the four clusters that do not host the superluminous disk galaxy at their cores and established that the centrals are massive elliptical galaxies. However, for two of the clusters, the offset superluminous spirals are brighter than the central galaxies, implying that the superluminous disk galaxies are the \textit{brightest} cluster galaxies. Our results demonstrate that superluminous disk galaxies are rarely the central systems of galaxy clusters. This is likely because galactic disks are destroyed by major mergers, which are more frequent in high-density environments. We speculate that the disks of superluminous disk galaxies in cluster cores may have been re-formed due to mergers with a gas-rich satellite. \\
\end{abstract}


\section{Introduction}
\label{sec:intro}

Brightest cluster galaxies (BCGs) are the most luminous and largest galaxies that reside in the highest density regions of the universe, most notably in the center of galaxy clusters. The large mass, the central location, and the extended stellar envelope of BCGs hint that they were formed by a series of accretion and merger events \citep{delucia07,lavoie16}. Theoretical studies also suggest that BCGs undergo a series of dissipationless mergers, which are mergers that destroy disks and result in the formation of a massive cD or elliptical galaxy \citep{delucia06,collins09}. Additionally, it is believed that dissipationless mergers play a crucial role in the assembly of the brightest cluster galaxies, which, in turn, result in the formation of an elliptical galaxy \citep{linden07,cattaneo11}. In agreement with this, observational studies demonstrate that BCGs are unequivocally galaxies with cD or elliptical morphology.

\begin{table*}
\caption{Characteristics of the candidate superluminous disk galaxies and the clusters in their vicinity}
\begin{minipage}{19cm}
\renewcommand{\arraystretch}{1.4}
\scriptsize	
\resizebox{0.95\textwidth}{!}{%
\begin{tabular}{cccccccccccc}
\hline
Name &  $z_{\rm gal}$ & $1\arcmin$  &  $N_{\rm H} $ & $M_{\rm \star}$ & SFR & $kT$  &  $R_{\rm 500}$ &   $L_{\rm 500}$ & \multicolumn{3}{c}{Separation}  \\
           & & (kpc) &($\rm{10^{20} \ cm^{-2}}$) & ($10^{11} \ M_{\rm \odot}$) & ($\rm{M_{\rm \odot} \ yr^{-1}}$) &(keV) & (kpc)&($10^{44} \ \rm{erg \ s^{-1}}$) & ($\arcsec$) & (kpc)& $1/R_{\rm 500}$ \\
(1) & (2) & (3) &(4) & (5) & (6) & (7) &(8) & (9) & (10)& (11) & (12) \\
\hline
2MASX J16273931+3002239 & $0.2599$ & $241.4$ & $2.0$ & $5.8$& 3.5 & $3.55^{+0.18}_{-0.20}$ & $797$ &  $1.17\pm0.12 $   & $2.4$ & $9.6$ & $0.01$ \\
2MASX J14475296+1447030 & $0.2207$ & $213.7 $& $1.7$ & $4.5$& 7.8 & $4.69^{+0.21}_{-0.19}$ & $953$ &  $3.13\pm0.14$   & $107.2$ & $378$ & $0.40$ \\
2MASX J09254889+0745051 & $0.1723$ & $175.8$ & $4.1$ & $4.2$& 7.1 & $2.35^{+0.18}_{-0.16}$ & $659$ &  $0.40\pm0.04$   & $111.4$ & $323$ & $0.49$ \\
2MASX J09572689+4918571 & $0.2414$ & $228.6$ &$0.9$ & $6.5$& 4.0 & $2.50^{+0.11}_{-0.10}$ & $661$ &  $1.13\pm0.16$   & $162.1$ & $612$ & $0.93$ \\
2MASX J14072225+1352512 & $0.2937$ & $263.3$ &$1.5$ & $8.3$& 3.0 & $3.30^{+0.25}_{-0.19}$ & $745$ &  $3.14\pm0.28$   & $564.6$ & $2459$ & $3.30$ \\
SDSS J113800.88+521303.9 & $0.2959$ & $264.7$ &$1.3$  & $3.9$& 6.2 & ... & ... & $<0.29$   & ... & ... & ... \\
SDSS J093540.34+565323.8 & $0.2964$ & $265.0$ &$2.4$  & $9.5$& ... & ... & ... & $<0.10$  & ... & ... & ... \\
\hline
\end{tabular}}
\vspace{0.15in}
\end{minipage}
Columns are as follows. (1) Identifier of the candidate superluminous disk BCG; (2) Redshift of the galaxy from SDSS DR9; (3) Physical scale at the redshift of the galaxy; (4) Total line-of-sight column  density (i.e.\ atomic and molecular hydrogen) based on the LAB survey \citep{willingale13}; (5) Stellar mass of the galaxy; (6) Star formation rate inferred from \textit{WISE} $12\ \rm{\mu m}$ luminosity \citep{ogle19}; (7) Best-fit gas temperature of the host (or nearby) galaxy cluster computed within the $R_{\rm 500}$ radius;  (8)  $R_{\rm 500}$ radius of the galaxy cluster; (9) $0.1-2.4$~keV band luminosity of the galaxy cluster within the $R_{\rm 500} $ radius; (10), (11), and (12) The projected distance between the candidate BCG and the center of the galaxy cluster in units of arc second, kpc, and $R_{\rm 500}$ radius, respectively. 
\label{tab:clusters}
\end{table*}

 \begin{table*}[!t]
\caption{The list of \textit{XMM-Newton} observations of superluminous disk galaxies}
\begin{minipage}{18cm}
\renewcommand{\arraystretch}{1.4}
\centering
\begin{tabular}{c c c c c }
\hline 
Galaxy & Obs ID & $t_{\rm total}$  & $t_{\rm clean}^{\dagger}$ & Date \\
name &  & (ks)  & (ks)& \\
\hline
2MASX J16273931+3002239  & 0842080501$^{\rm{\ddagger}}$ & 22.0 &$-$& 2019 Jul 22 \\
         & 0861610101$^{\rm{\ddagger}}$ & 24.0 &$-$& 2020 Aug 21 \\
        & 0861610201 & 28.0 &20.9/22.0/11.9& 2021 Jan 17 \\
2MASX J14475296+1447030  & 0842080201 & 13.0 &11.0/11.1/11.7& 2019 Jul 21 \\
2MASX J09254889+0745051  & 0842080301 & 11.0 &9.2/9.4/6.3& 2019 May 16  \\
2MASX J09572689+4918571  & 0842080801 & 19.7 &17.2/17.5/13.7& 2019 Apr 29 \\
2MASX J14072225+1352512  & 0842080401 & 22.4 &17.7/20.1/9.0& 2019 Jun 28\\
SDSS J113800.88+521303.9  & 0842080601 & 36.9 &3.1/4.2/2.5& 2019 Dec 01 \\
SDSS J093540.34+565323.8  & 0842080701 & 34.0 &26.1/26.6/18.7& 2019 May 13 \\
 \hline \\
\end{tabular} 
\end{minipage}
$^{\dagger}$ The clean exposure times refer for the EPIC PN, MOS1, and MOS2 cameras, respectively.\\
$^{\ddagger}$ Data from these observations were not used for the analysis due to the high background level and technical problems associated with the observations (see Section \ref{sec:data} for details). 
\vspace{0.5cm}
\label{tab:xmmdata}
\end{table*}

Similar to other massive galaxies, BCGs also host black holes (BHs), which are among the most massive BHs in the universe \citep{mcconnell12}. The mass of these BHs is believed to correlate with the total mass of the galaxy cluster with more massive clusters hosting more massive BHs in their BCGs \citep{bogdan18,phipps19,lakhchaura19}. The BHs in these BCGs can shine as  active galactic nuclei (AGN) and can release copious amounts of energy. This energy input  heats the gas and can either expel it to large radii or completely eject it from the large-scale dark matter halo. Because this energetic feedback offsets gas cooling, and, hence, the star formation in BCGs, these galaxies are expected to be quiescent. Indeed, observational studies demonstrated that BCGs have negligible star-formation \citep{cluver14}, except for some extreme cooling flow systems, such as the Phoenix cluster \citep{mcdonald12}. 

In contrast with the above-described picture, a population of disk-dominated galaxies, so-called ``super spiral'' and ``super lenticular'' galaxies, were discovered based on optical (Sloan Digital Sky Survey  -- SDSS) and infrared (Wide-field Infrared Survey Explorer  -- WISE) data \citep{ogle16,ogle19}. Many physical properties of these galaxies are comparable with BCGs; they are among the most massive galaxies in the universe with $M= (3-10)\times10^{11} \ \rm{M\odot}$ and are very extended with $D=57-134$ kpc. These galaxies also exhibit substantial star-formation with their star-formation rate being in the range of $ 3-65 \ \rm{M_{\odot} \ \rm{yr^{-1}}}$. \citet{ogle16} suggested that a sub-sample of super spiral and super lenticular galaxies may reside in rich galaxy cluster environments and some of them could even be the BCG of their host cluster. To identify these candidate BCGs, \citet{ogle16} searched for previously identified galaxy clusters in the proximity ($\Delta r < 1\arcmin$) of the super spiral and super lenticular galaxies that also have small redshift difference ($\Delta z <0.04 $) relative to the cluster. These conditions suggest that the super spiral and lenticular galaxies are physically associated with the galaxy clusters and are not simply projected on the clusters. However, it is important to emphasize that the identification of BCGs based on optical data can lead to misclassification because it is difficult to identify the cluster centers \citep{miller05,linden07}. Therefore, while results based on SDSS are indicative, they cannot conclusively identify the BCGs of galaxy clusters. 

X-ray observations of the intracluster medium (ICM) of galaxy clusters offer a robust means to determine the position of cluster centers. In relaxed systems, BCGs reside at the bottom of the galaxy cluster's potential well, which coincides with the peak of the X-ray emission from the ICM. Based on a study of 62 galaxy clusters, \citet{zhang11} demonstrated that the typical offset between the peak of the X-ray emission and the BCG is $\sim10$~kpc. Given that most super spiral and super lenticular galaxies reside at a redshift of $z=0.2-0.3$, this offset corresponds to a projected distance of $\Delta r = 2.3\arcsec-3.1\arcsec$, which is much smaller than the search radius used by \citet{ogle16}. Therefore, we initiated an X-ray observing campaign to systematically probe whether superluminous disk galaxies reside in the center of galaxy clusters. We collected \textit{XMM-Newton} X-ray data of the galaxy clusters whose BCG could be a superluminous disk galaxy. Our initial results of five galaxy clusters were published in \citet{bogdan18b}, where we identified a superluminous lenticular galaxy as a BCG. In this work, we follow-up on this study and complement our galaxy cluster sample with seven additional objects. 

This paper is structured as follows. We describe the sample of superluminous disk galaxies in Section 2. The \textit{XMM-Newton} data and the reduction procedure are discussed in Section 3. In Section 4 we present our results, including the determination of cluster centroids and probing whether the superluminous disk galaxies are the BCGs of the clusters. We discuss our results in Section 5 and summarize in Section 6. Throughout the paper we assume $H_0=70 \ \rm{km \ s^{-1} \ Mpc^{-1}}$, $ \Omega_M=0.3$, and $\Omega_{\Lambda}=0.7$, and all error bars are $1\sigma$ uncertainties.

\begin{figure*}[!t]
  \begin{center}
    \leavevmode
      \epsfxsize=0.33\textwidth \epsfbox{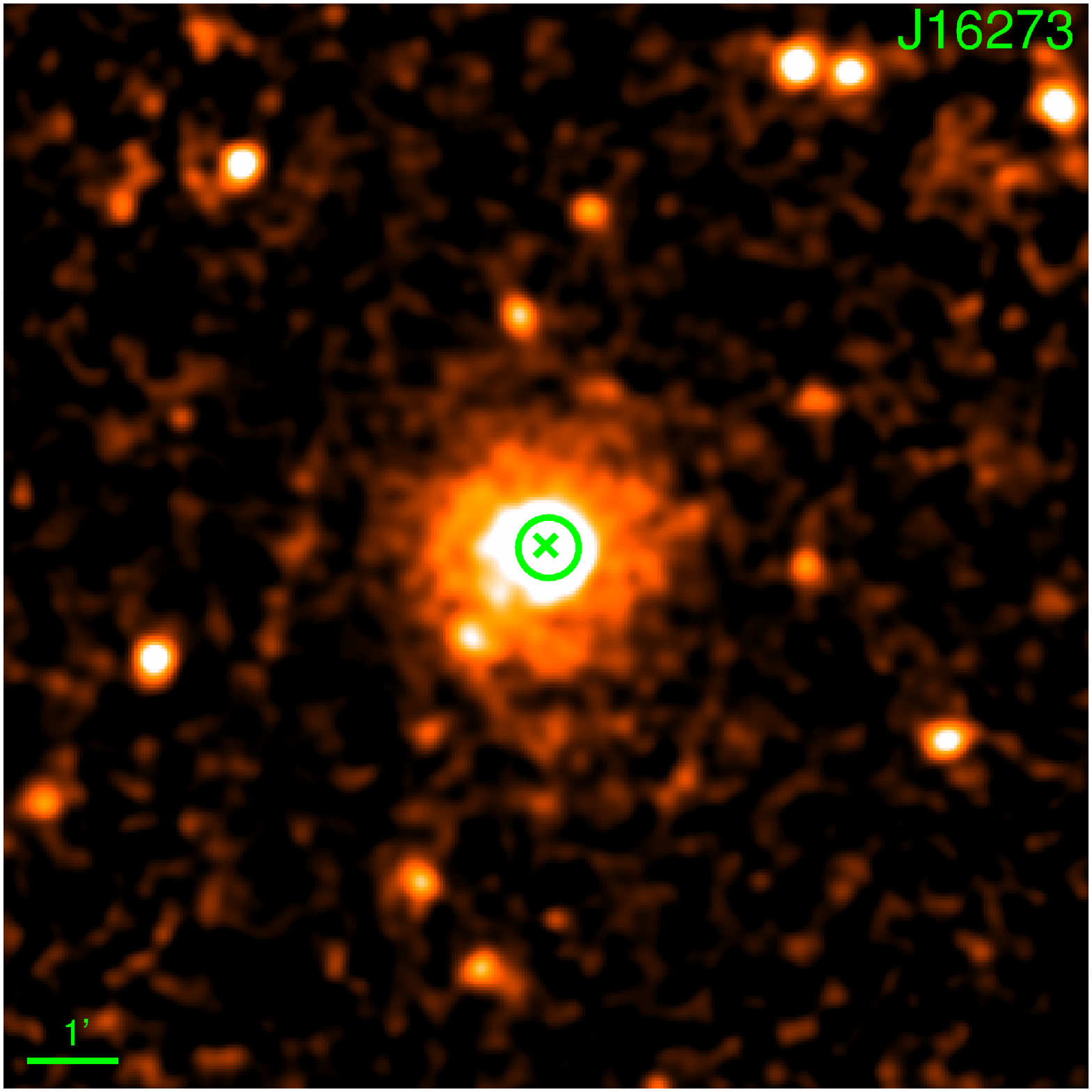}
      \epsfxsize=0.33\textwidth \epsfbox{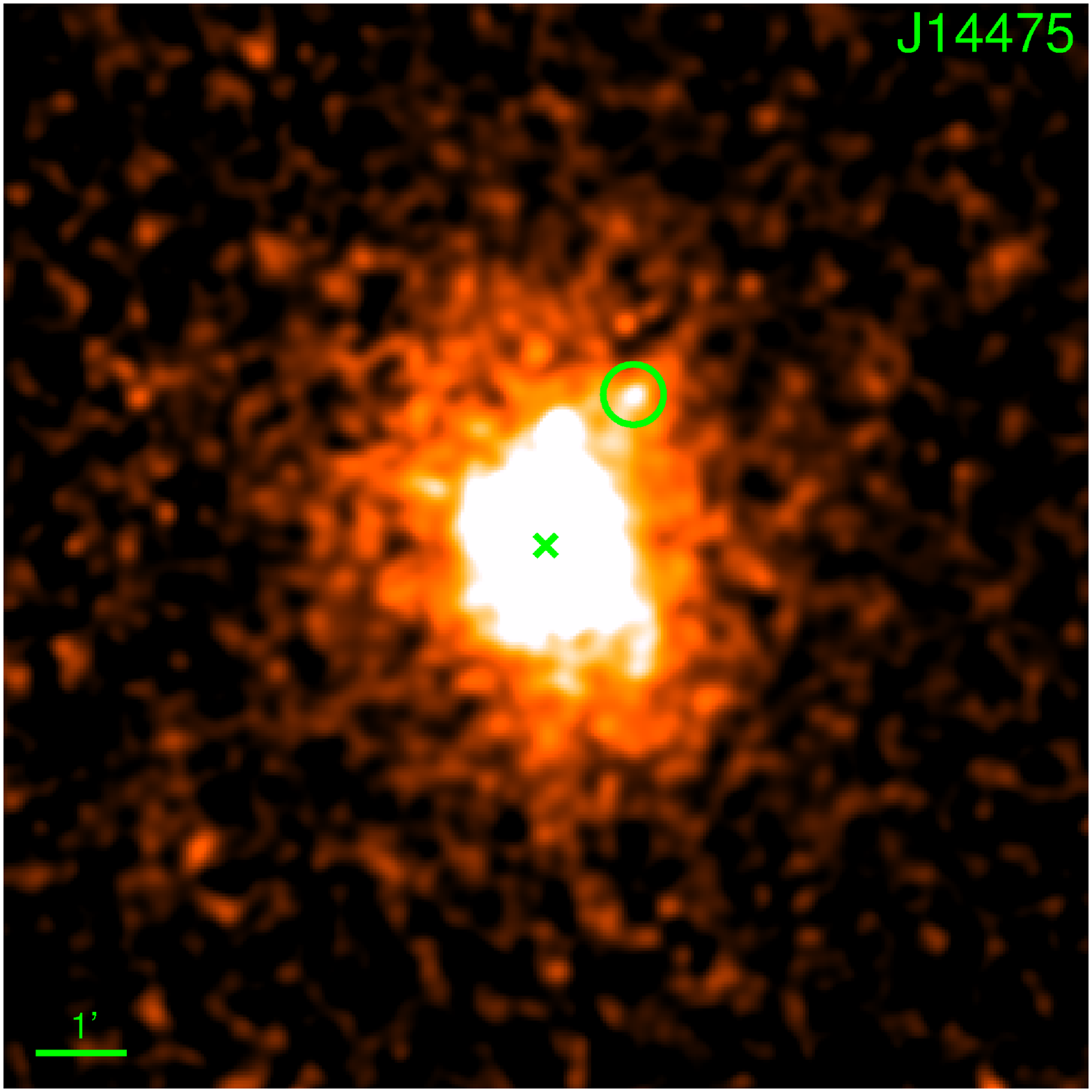}
      \epsfxsize=0.33\textwidth \epsfbox{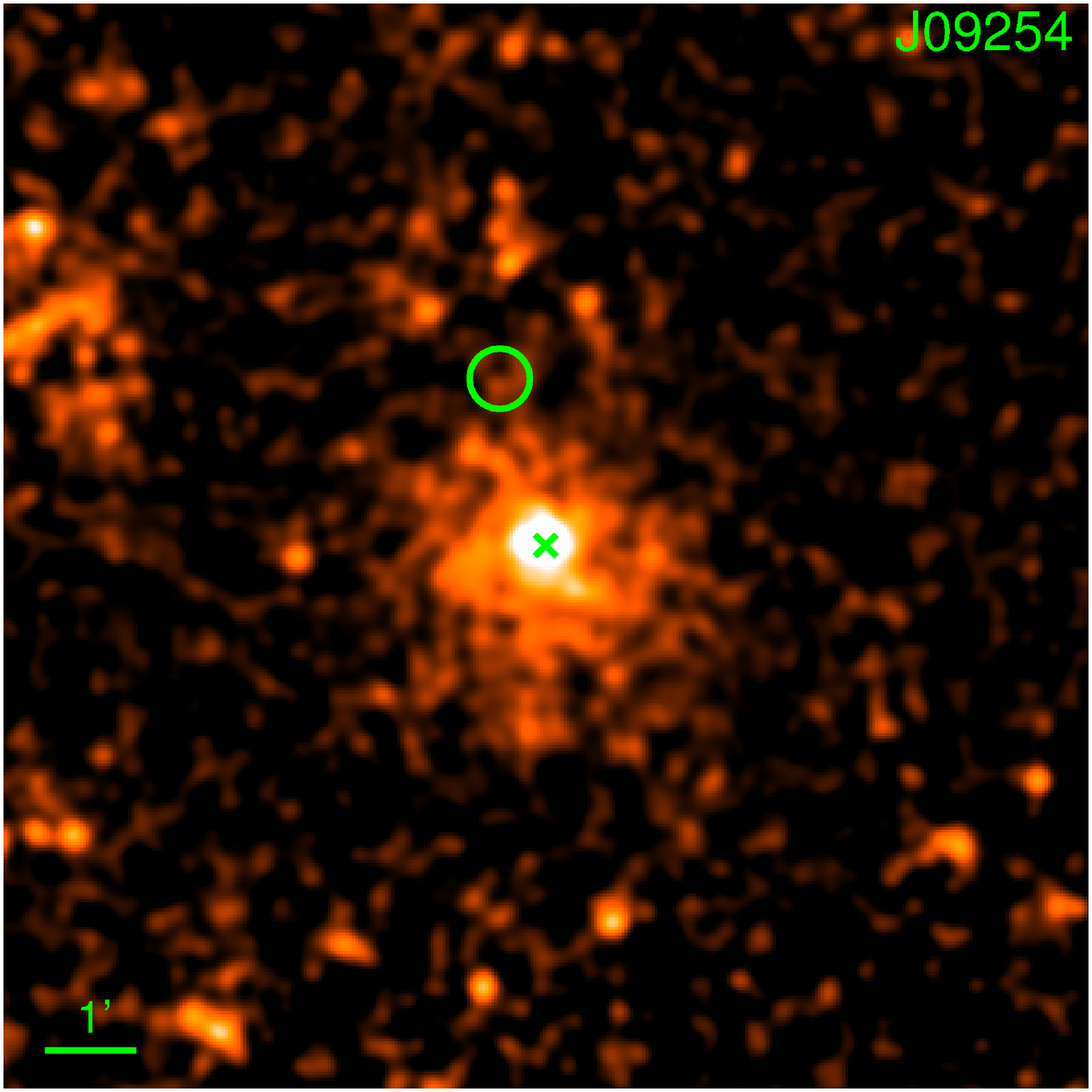}
      \epsfxsize=0.33\textwidth \epsfbox{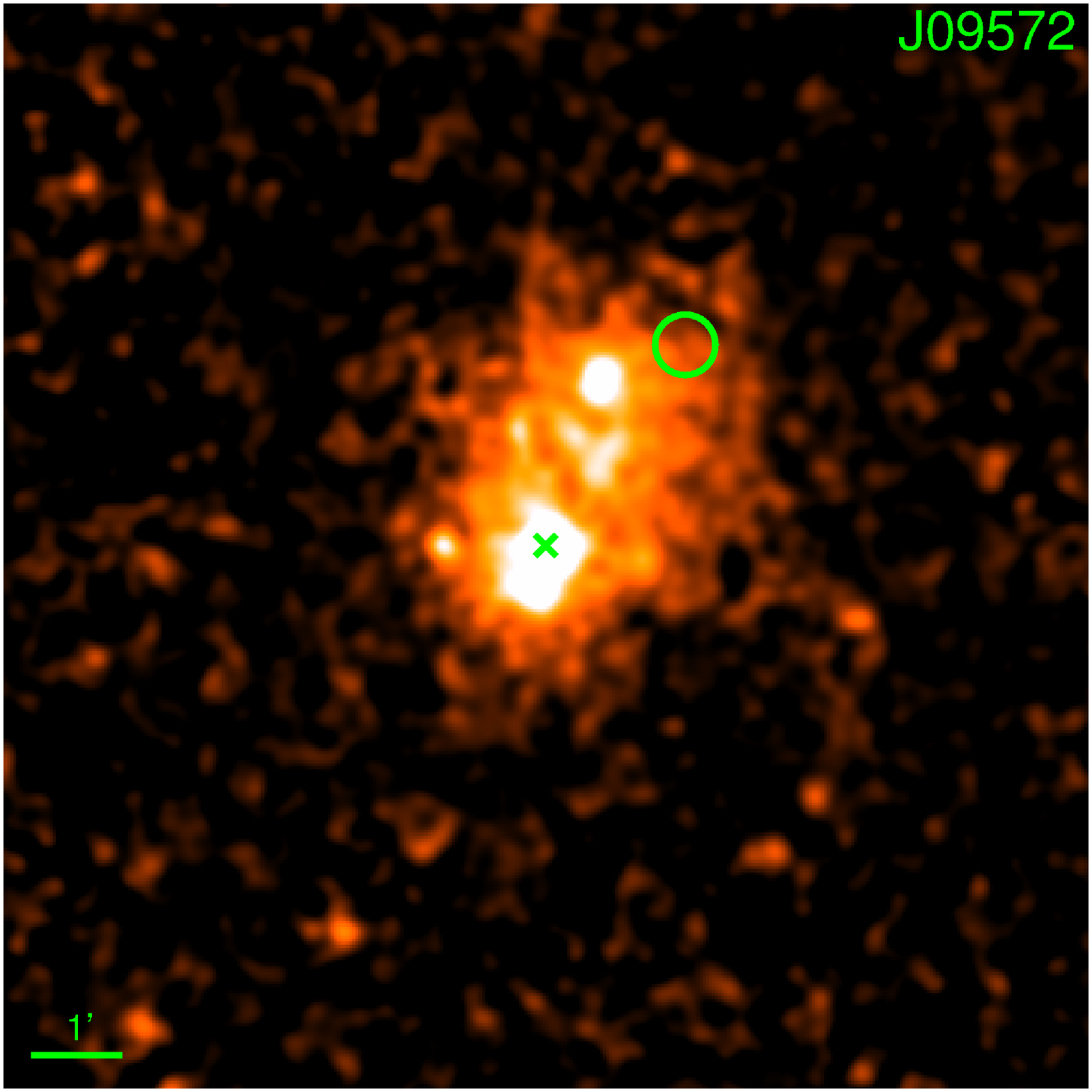}
      \epsfxsize=0.33\textwidth \epsfbox{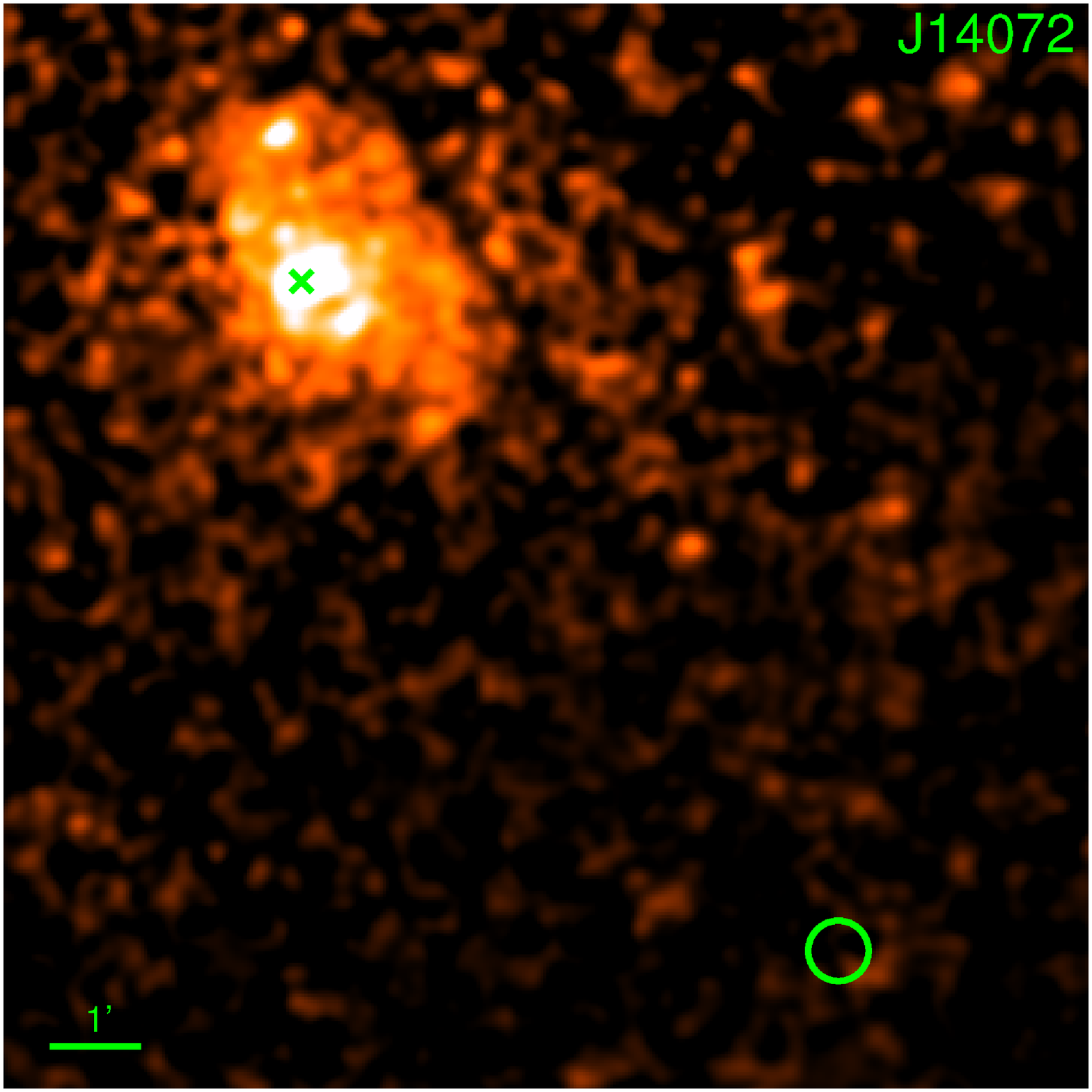}
      \epsfxsize=0.33\textwidth \epsfbox{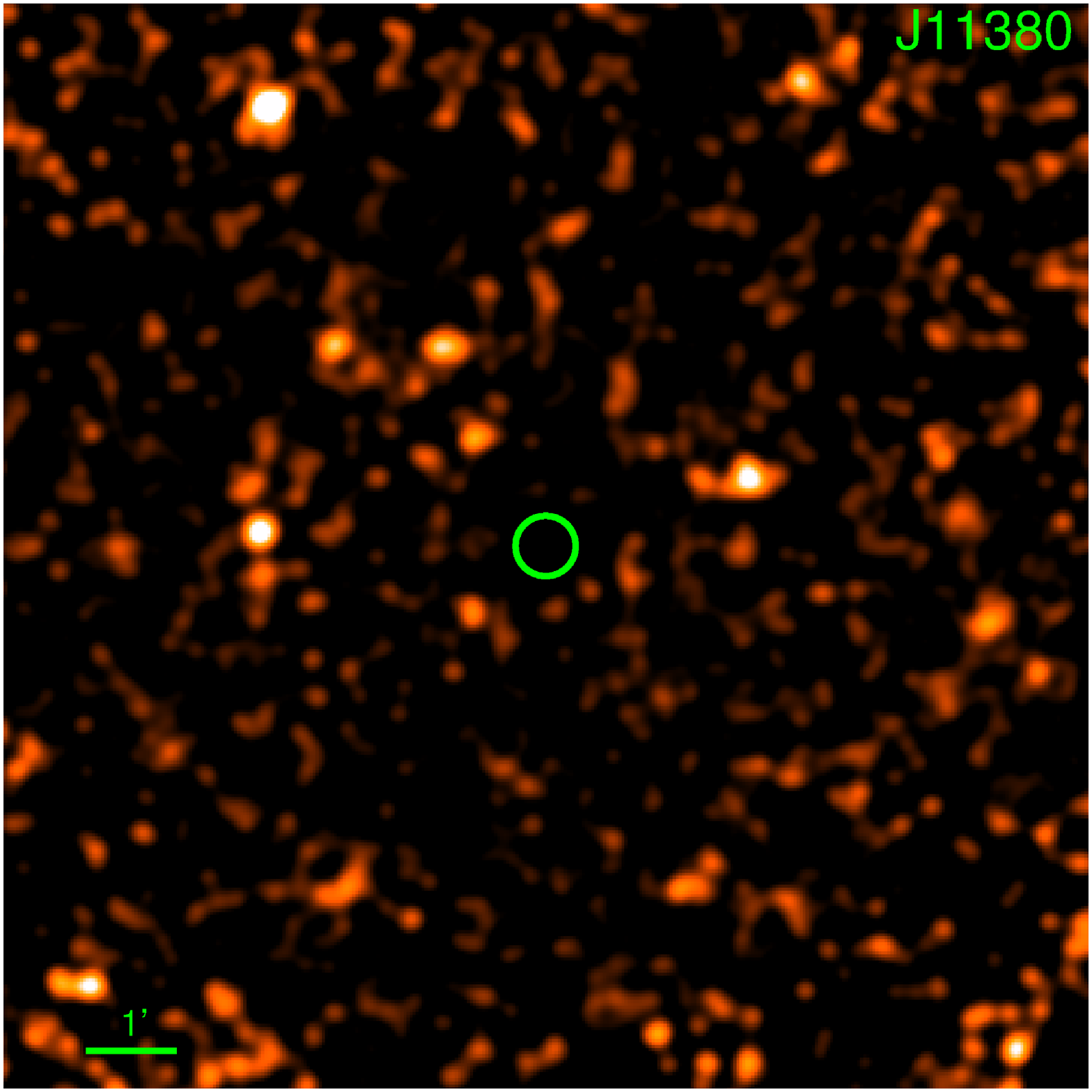}
      \epsfxsize=0.33\textwidth \epsfbox{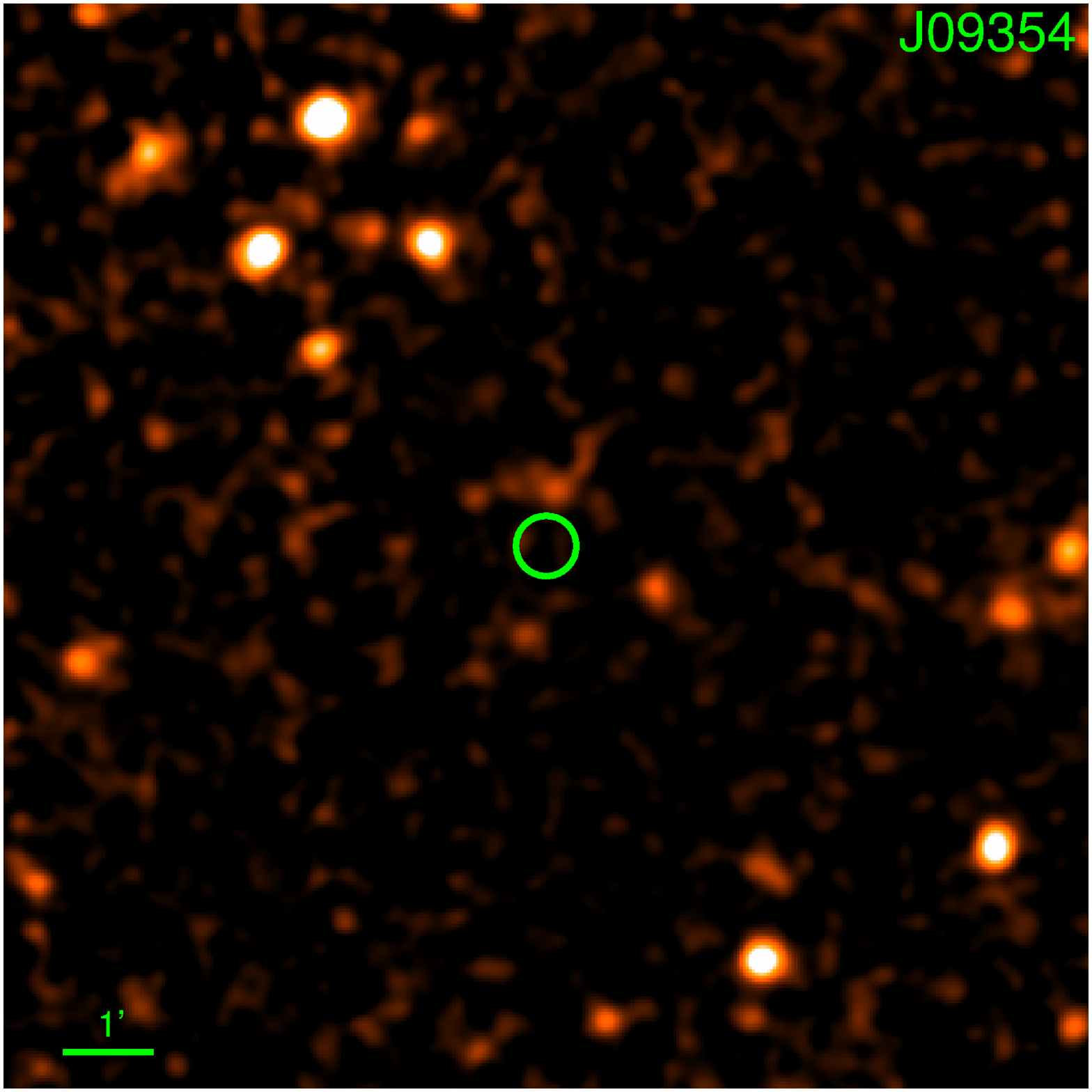}
      \vspace{0cm}
      \caption{$0.7-2$~keV band \textit{XMM-Newton} X-ray images of the superluminous disk galaxies and their large-scale environment. Data from the EPIC PN and EPIC MOS cameras are co-added. The images are background subtracted and exposure corrected. The circles show the position of the superluminous disk galaxies and the cross marks the peak of the ICM emission, which represents the center of the galaxy clusters. We identified one superluminous spiral galaxy, which is located at the center of a galaxy cluster (top left panel). Note that we did not detect extended X-ray emission (i.e.\ galaxy groups or clusters) in the vicinity of two superluminous disk galaxies (bottom panels).} 
     \label{fig:xmmimages}
  \end{center}
\end{figure*}

\begin{figure*}[!htp]
  \begin{center}
    \leavevmode
      \epsfxsize=0.7\textwidth \epsfbox{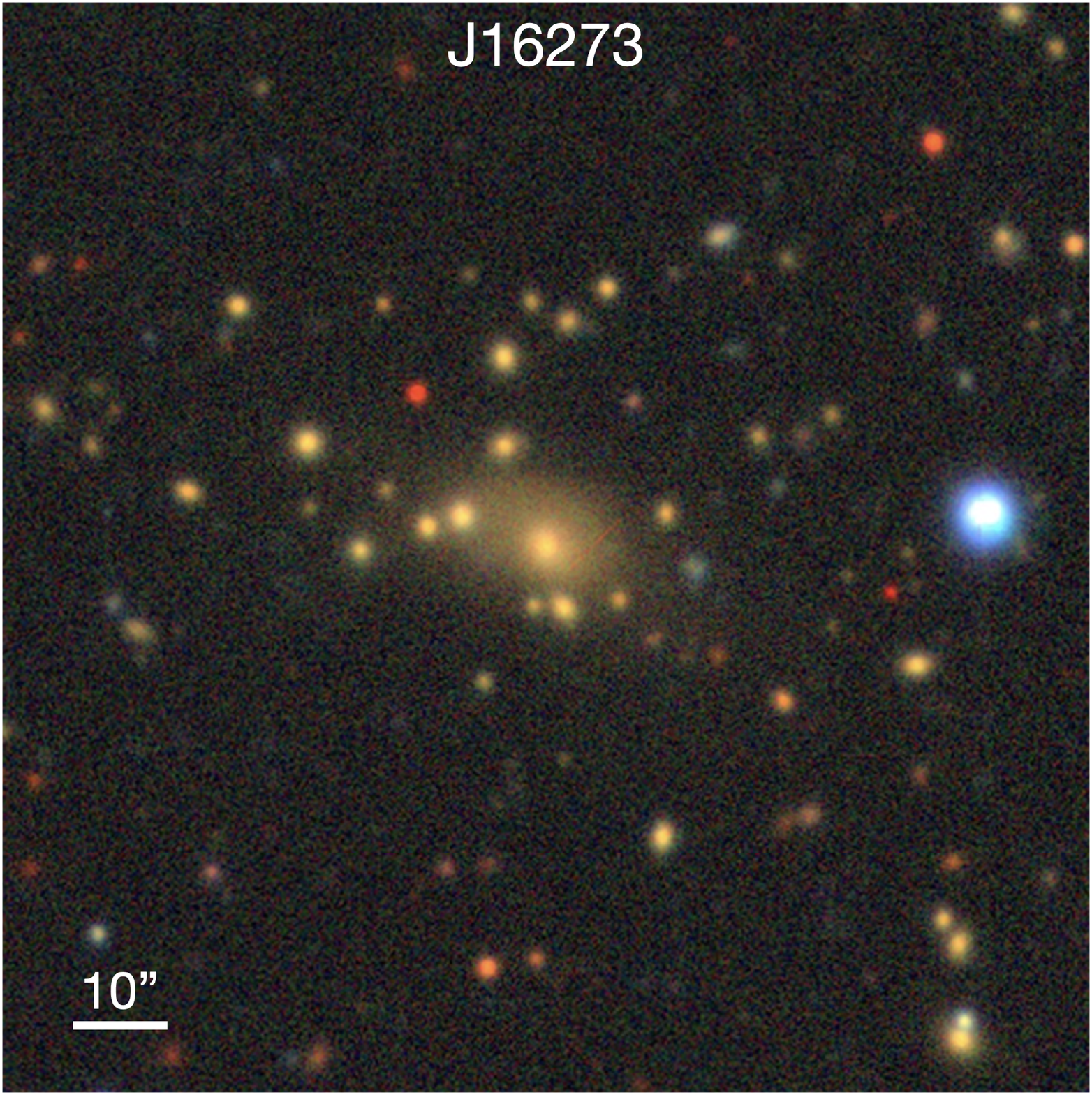}
      \vspace{0cm}
      \caption{DESY Legacy Survey image of the superluminous spiral galaxy, J16273, and its environment. This galaxy resides in the core of a massive galaxy cluster. The side length of the image is $115\arcsec \times 115\arcsec$ ($460\times460$~kpc). The galaxy exhibits a lopsided disk, which may hint at a past merger. There are a number of low-mass galaxies in the vicinity of J16273. The offset between the galaxy and the cluster center is $2.4\arcsec$, which corresponds to $9.6$~kpc.}
     \label{fig:spiralbcg}
  \end{center}
\end{figure*}

\section{The sample of superluminous disk galaxies}
\label{sec:sample}

To identify our targets, we relied on the catalog of superluminous disk galaxies provided by \citet{ogle16,ogle19}; these works identified  28 superluminous disk galaxies that could be members of galaxy groups or clusters. In \citet{bogdan18b}, we studied five galaxy clusters and determined that one superluminous lenticular is associated with a low-mass galaxy cluster. While our earlier work was based on the initial sample of \citet{ogle16}, here we rely on the sample of \citet{ogle19} to identify other superluminous disk galaxies that are candidate BCGs. 

The X-ray emission from the ICM can robustly identify massive galaxy clusters. To this end, we utilized X-ray observations from the \textit{ROSAT} All-Sky Survey and searched for statistically significant X-ray emission that may originate from the large-scale ICM. We found 7 new superluminous disk galaxies that exhibit $\gtrsim2\sigma$ detection at the approximate position of the superluminous disk galaxies. Assuming that the hot gas originates from thermal plasma with $kT=2$~keV and a metallicity of $Z=0.3 \ \rm{Z_{\odot}}$, we estimated that the luminosity of the clusters is $(0.5-3)\times10^{44} \ \rm{erg \ s^{-1}} $. These luminosities are typical for massive galaxy clusters. The available \textit{ROSAT} data of these clusters is rather shallow hence the ICM properties cannot be characterized. Additionally, the ROSAT PSPC detector does not have sufficiently high angular resolution to determine the centroid of the ICM emission. Therefore, we collected \textit{XMM-Newton} observations of these seven systems. For the other galaxy clusters, we did not detect statistically significant emission, implying the lack of luminous ICM emission. The properties of the candidate BCGs and the galaxy clusters are listed in Table \ref{tab:clusters}. The derived physical properties of the galaxy clusters are discussed in Section \ref{sec:properties}.

\section{\textit{XMM-Newton} data analysis}
\label{sec:data}

In this work, we analyze \textit{XMM-Newton} X-ray observations of seven superluminous disk galaxies and their surroundings. The observations were taken in AO-18 except for 2MASX\,J16273931+3002239 (henceforward J16273), which was also observed in AO-19 because the original observations were severely contaminated with high background time periods. We list the analyzed observations in Table  \ref{tab:xmmdata}. All data were taken with the European Photon Imaging Camera (EPIC). The analysis was carried out using the XMM Science Analysis System (SAS) version 19.0 and Current Calibration Files (CCF) as present in September 2021.

\begin{figure*}[!htp]
  \begin{center}
    \leavevmode
      \epsfxsize=0.35\textwidth\epsfbox{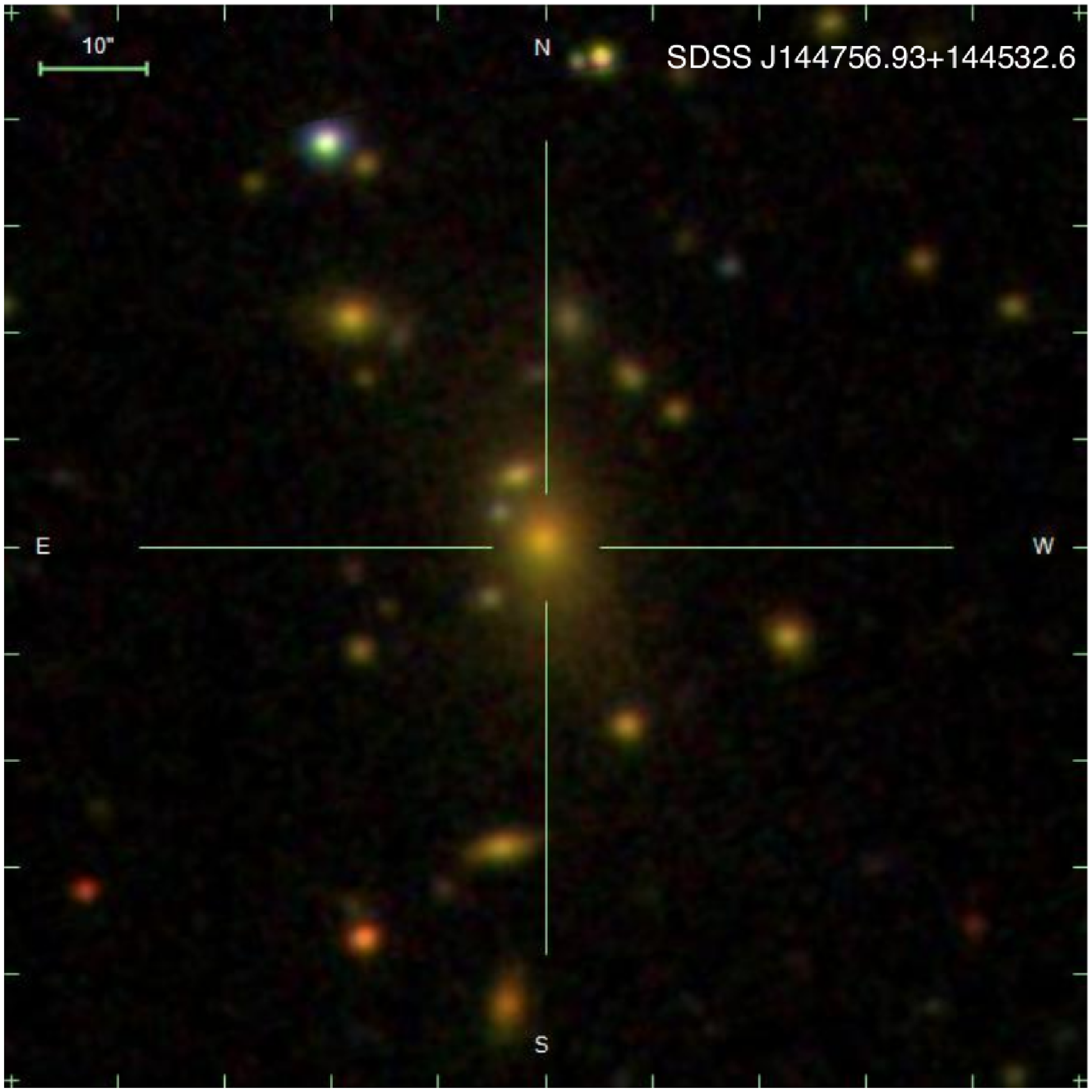}
      \epsfxsize=0.35\textwidth \epsfbox{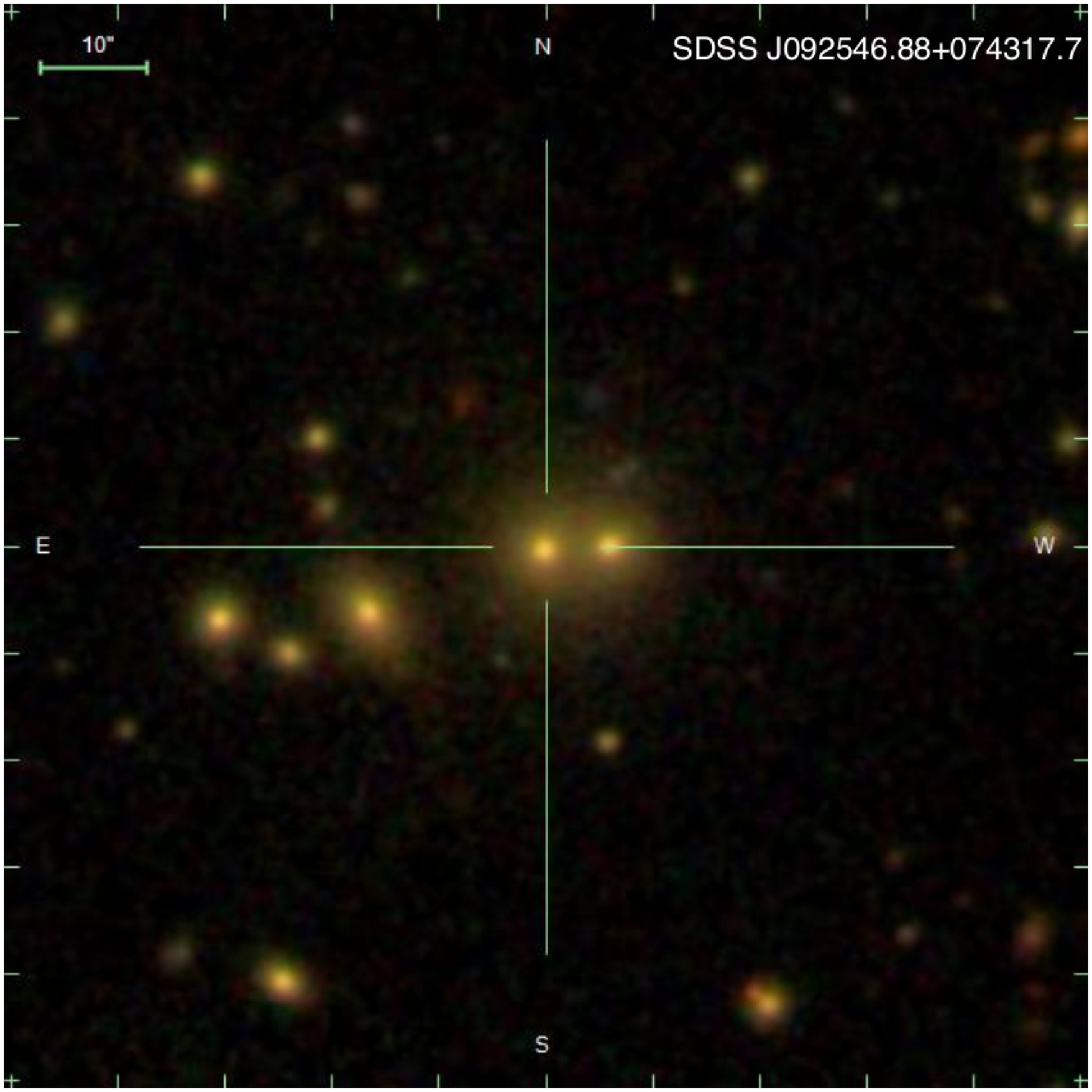}
      \epsfxsize=0.35\textwidth \epsfbox{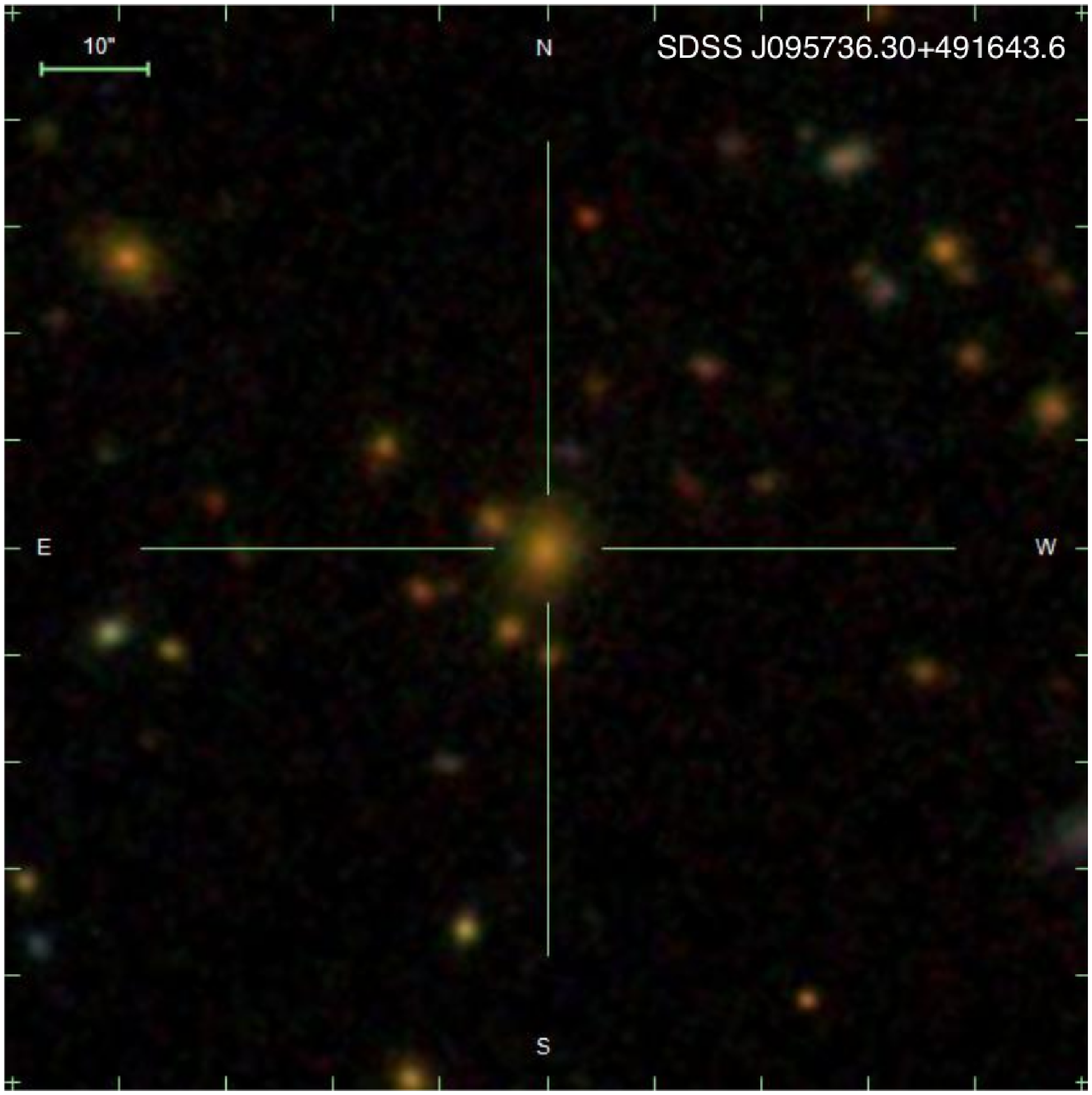}
      \epsfxsize=0.35\textwidth \epsfbox{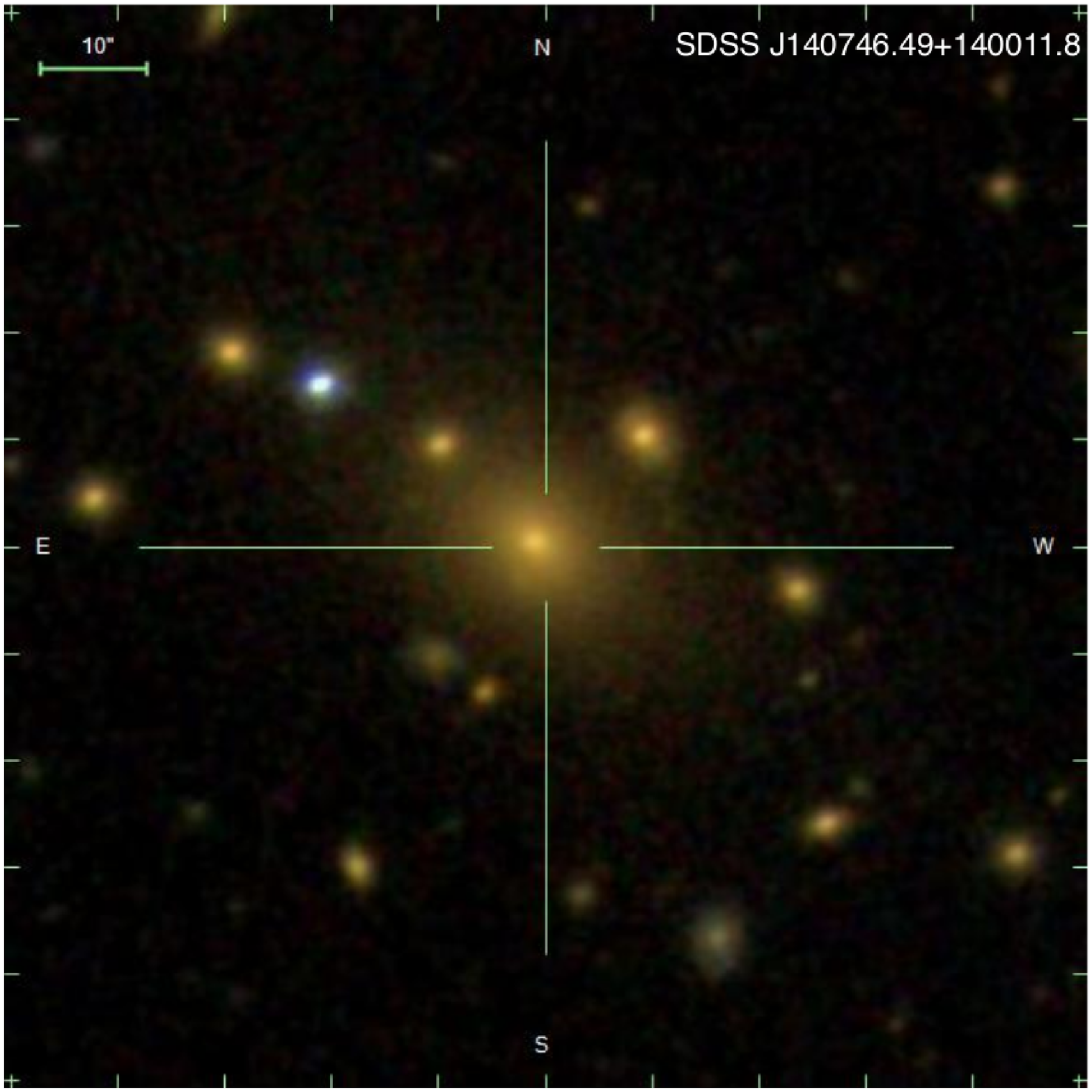}
      \vspace{0.5cm}
      \caption{SDSS composite (\textit{u,g,r,i,z}) images of the four galaxies residing in the center of the galaxy clusters, which do not host a superluminous disk galaxy at their heart. These galaxies exhibit elliptical morphology and their color indices also suggest that they are early-type systems.} 
     \label{fig:realbcg}
  \end{center}
\end{figure*}

The main steps of the data analysis agree with those outlined in previous works \citep{lovisari19,lovisari20}. We first generated calibrated event files for the EPIC-PN and EPIC-MOS data using the \textit{epchain} and \textit{emchain} tasks and included events with patterns $\leq4$ and $\leq12$, respectively. 

Because the detectors aboard \textit{XMM-Newton} are sensitive to Solar flares, it is essential to identify and exclude high background periods. We discarded the data corresponding to the periods of high background using the tasks {\it mos-filter} and {\it pn-filter}. We note that ObsID:~0842080501 was very heavily contaminated with flares with $93\%$ and $73\%$ of the EPIC-PN and EPIC-MOS time contaminated, hence the science data was not used in the present analysis. This motivated us to re-observe this target, which was carried out in ObsID:~0861610101. However, due to technical problems, these data do not include EPIC-MOS observations, hence the target was observed again in ObsIDs:~0861610201, which observations provide data for all detectors. The original and clean exposure times are tabulated in Table \ref{tab:xmmdata}. 

Because we are primarily interested in the diffuse emission associated with galaxy clusters, we must identify bright point sources, which are mostly associated with high-redshift AGN. To this end, we utilized the \textit{edetect\_chain} task. This tool generates a source list using multiple energy ranges, which was used to mask the point sources from the analysis of the diffuse emission. 

To account for the instrumental and sky background components, we refer the reader to \citet{lovisari19}, which background subtraction method was followed in the present analysis.

\section{Results}
\label{sec:results}

\subsection{XMM-Newton X-ray Images}
\label{sec:images}

In Figure \ref{fig:xmmimages}, we present the $0.7-2$~keV \textit{XMM-Newton} images of the superluminous disk galaxies and their large-scale environment. The X-ray images are background subtracted and exposure corrected and include data from both the EPIC-PN and EPIC-MOS detectors. Five X-ray images reveal large-scale extended X-ray emission, which originates from massive galaxy clusters. 
The X-ray images associated with SDSS~J113800.88+521303.9 and SDSS~J093540.34+565323.8 do not show extended emission within the studied regions. This implies that these galaxies do not reside in the vicinity of galaxy clusters. While the \textit{ROSAT} X-ray images showed excess emission associated with these galaxies, this emission was likely associated with bright point sources (e.g.\ AGN) that were resolved and removed from the \textit{XMM-Newton} images. Because the detection significance in the \textit{ROSAT} images was rather low, it is also possible that the excess X-ray emission in the \textit{ROSAT} data is due to an upward fluctuation. The properties of the detected galaxy clusters are discussed in Section \ref{sec:properties}.

\subsection{Offset from the Cluster Centers}
\label{sec:offset}

To determine the centroid of the galaxy clusters, we utilize the background-subtracted X-ray images (Figure \ref{fig:xmmimages}) and apply  Gaussian smoothing with a kernel size of $6\arcsec$. The peak of the X-ray emission on the smoothed images defines the cluster center. We then measure the offset between the position of the superluminous disk galaxies and the cluster centers. This offset is the projected distance between the superluminous disk galaxies and the center of the cluster's gravitational potential. 

For one galaxy, J16273, the projected distance from the cluster center is only $2.4\arcsec$, which corresponds to $9.6$~kpc at the redshift of the galaxy ($z=0.2599$). This offset is comparable to the typical offset between the X-ray peak of galaxy clusters and the position of BCGs \citep{zhang11}. Additionally, this small offset may also be caused by projection effects. The Legacy Survey image of the galaxy is presented in Figure \ref{fig:spiralbcg}. Visual inspection of this image indicates that all galaxies in the proximity of J16273 are less luminous. To qualitatively study the neighboring galaxies around J16273, we performed an environment search using the NASA/IPAC Extragalactic Database. We conservatively searched for galaxies within the redshift range of $z=0.23-0.29$ considering the uncertainty of the photometric redshift of the neighboring galaxies. We identified six galaxies in this redshift range within a projected distance of $0.6\arcmin$ ($145$~kpc). By deriving the $r-$band luminosity of the galaxies, we found that the neighboring galaxies are factor of $4-630$ less luminous than J16273. This implies that J16273 is the BCG of this galaxy cluster.


The offset between the superluminous disk galaxies and the centroid of galaxy clusters is much larger for the remaining four systems. Specifically, for three clusters the offset ranges between $107.2\arcsec-162.1\arcsec$, which corresponds to a projected distance of $323-612$~kpc. These large offsets exclude the possibility that the superluminous disk galaxies are BCGs of the galaxy clusters. However, given that the redshift of the disk galaxies and the galaxy clusters are similar, and the projected distance is smaller than the $R_{\rm 500}$ radius of the galaxy clusters, it is likely that these superluminous disk galaxies are physically associated with the galaxy clusters. We note that 2MASX~J09572689+4918571 is associated with a merging galaxy cluster. Therefore, it cannot be excluded that this superluminous disk galaxy was the central galaxy of the northeastern sub-cluster, whose core is located at a projected distance of $60\arcsec$ ($\approx223$~kpc) from the superluminous disk galaxy. For the remaining one system, 2MASX\,J14072225+1352512, the offset is $564.6\arcsec$, which represents a projected distance of $2459$~kpc. This offset is $\approx3.3$ times larger than the $R_{\rm 500}$ of the galaxy cluster, therefore it is unlikely that this galaxy belongs to the cluster. Instead, this galaxy is likely projected onto the same region of the sky.

\subsection{Superluminous Spiral at the Center of a Galaxy Cluster}
\label{sec:galaxy_center}

To explore the physical properties of J16273, we reanalyzed its SDSS optical and WISE infrared data.  The global stellar mass and star formation rate are estimated from the WISE mid-infrared photometry, where the short wavelength bands of WISE are sensitive to the host stellar component and the longer wavelengths the star formation activity \citep{jarrett13}.  Based on the calibration of \citet{cluver14}, the $W1-W2$ color may be used to estimate the stellar mass-to-light,  and the light derives from the W1 in-band luminosity.  The stellar mass hence follows, with uncertainties ($10-20\%$) propagating from the calibration and WISE photometry.  For the star formation rate, we use the calibration from \citet{cluver17}, where the mid-infrared luminosity ($W3$ and $W4$ bands) is tightly correlated to the total-infrared luminosity.  Typical uncertainties, from the calibration and photometric measurements, are $20 -40\%$.  Using the color-color relations in \citet{jarrett19}, we note that the WISE colors,  $W1-W2 = 0.01 \pm 0.05$ and $W2-W3 = 1.42 \pm 0.28$ [Vega mag],  essentially relating the stellar population to the SF activity, have values that indicate a transition between spheroid-dominated (e.g., tri-axial early types) and intermediate disk galaxies (disk + bulge).  The colors are consistent with the relatively intermediate-to-large bulge-to-total  - $B/T = 0.49$ — that is inferred from the W1 radial profile using the decomposition detailed in \citet{jarrett19}.

We also inspected the optical spectrum of the galaxy to search for potential AGN signatures. To this end, we utilized the SDSS DR12 optical spectrum of the galaxy and computed the standard optical diagnostic line ratios, [O~III]/H$\rm{\beta}$, [N~II]/H$\rm{\alpha}$, and [S~II]/H$\rm \alpha$, which allow us to place the galaxy on the [NII]-BPT and [SII]-BPT diagrams \citep{baldwin81}. We obtained $\log \rm{([OIII]/H\rm{\beta})} = -0.83$, $\log \rm{[NII/H\rm{\alpha}]} = -0.38$, and $\log \rm{[SII/H\rm{\alpha}]} = -0.21$. These values place J16273 below the star-forming locus, with no indication of an AGN.

\subsection{Identifying the Central Galaxies of Four Clusters}
\label{sec:real_bcg}

In Section \ref{sec:offset} we demonstrated that there is only one superluminous spiral galaxy in our sample that resides in the center of a galaxy cluster. Because our X-ray analysis identified the center of each galaxy cluster, we can search for the true central galaxies in the other four clusters. To identify these galaxies, we investigated SDSS observations around the cores of the clusters and searched for the optically brightest galaxies. We identified massive galaxies in the center of each cluster, which are shown in Figure \ref{fig:realbcg}. 

To probe the characteristics of the true BCGs, we first measured their $u-r$ color indices, which hint at whether the galaxies are quiescent of star-forming. For the BCGs galaxies, we found $u-r = 3.18-4.01$, suggesting that these systems are red and quiescent \citep{strateva01}. Their morphologies also indicate that these are massive elliptical galaxies. Thus, the other galaxy clusters detected in the \textit{XMM-Newton} images host common massive ellipticals at their core. 

Interestingly, the brightness of the central elliptical galaxies is comparable to or even lower than the luminosity of the superluminous disk galaxies. The superluminous disk galaxy, 2MASX\,J14475296+1447030, is slightly fainter in $r$-band ($r_{\rm disk} - r_{\rm elliptical} = 0.37$~mag) but is brighter in the $u$-band ($u_{\rm disk} - u_{\rm elliptical} = -0.50$~mag) than the elliptical galaxy (SDSS\,J144756.93+144532.6) in the cluster core. Surprisingly, 2MASX\,J09254889+0745051 and 2MASX\,J09572689+4918571 are both brighter in the five SDSS bands than the elliptical galaxy in the cluster cores with $r_{\rm disk} - r_{\rm elliptical} = - 0.54$~mag and $r_{\rm disk} - r_{\rm elliptical} = - 1.07$~mag, respectively. Therefore, while the superluminous disk galaxies do not reside in the central regions of the galaxies, two of them (2MASX\,J09254889+0745051 and 2MASX\,J09572689+4918571) should still be considered as the \textit{brightest} cluster galaxy of their clusters.

 \begin{table}[!t]
\caption{The concentration and centroid-shift parameters of the galaxy clusters}
\begin{minipage}{8cm}
\renewcommand{\arraystretch}{1.4}
\centering
\begin{tabular}{c c c}
\hline 
Galaxy & c & w  \\
\hline
2MASX J16273931+3002239  & 0.206 & 0.0103 \\
2MASX J14475296+1447030  & 0.155 & 0.0064 \\
2MASX J09254889+0745051  & 0.126 & 0.0147  \\
2MASX J09572689+4918571  & 0.091 & 0.0792 \\
2MASX J14072225+1352512  & 0.072 & 0.0288 \\
 \hline \\
\end{tabular} 
\end{minipage}
\vspace{0.5cm}
\label{tab:morphology}
\end{table}

\subsection{Average Properties of the Galaxy Clusters}
\label{sec:properties}

Using the \textit{XMM-Newton} observations, we determined the properties of the galaxy clusters that were detected in the vicinity of the superluminous disk galaxies. To determine the best-fit temperature, X-ray luminosity, and total mass of the clusters within the $R_{\rm 500}$ region, we used an iterative process. Specifically, we computed the temperature in the region that has the highest signal-to-noise ratio. We then obtained the first estimate of the $R_{500}$ through the $kT-M_{\rm 500}$ relation established in \citet{lovisari20}. Then we extracted a second spectrum using the obtained $R_{\rm 500}$ to obtain a new cluster temperature. We continued this procedure until the $R_{\rm 500}$ converged, i.e.\ until the new $R_{\rm 500}$ did not differ more than $0.1\arcmin$ from the previous one. This region was used to measure the final best-fit temperature and luminosity. To determine the total mass of cluster, we relied on the $kT-M_{\rm 500}$ relation \citep{lovisari20}. 

To fit the spectra throughout this procedure, we used an APEC model, assumed Galactic column density \citep{willingale13} and used the metallicity tables of \citet{asplund09}. During the fitting process, the temperature, metallicity, and normalization were free parameters. The redshift of the cluster was fixed at the redshift of the superluminous disk galaxy. The best-fit ICM temperatures are in the range of $kT = 2.5-4.7$~keV. The $R_{\rm 500}$ radius of the clusters vary in the range of $R_{\rm 500} = 659 - 953$~kpc. We also derived the X-ray luminosity of the clusters within this radius and obtained $L_{\rm 0.1-2.4keV} = (0.40-3.14)\times 10^{44} \ \rm{erg \ s^{-1}}$. The best-fit temperatures, luminosities, and the inferred $R_{\rm 500}$ values are listed in Table \ref{tab:clusters}.

We did not detect extended X-ray emission aroud two superluminous disk galaxies. While this non-detection suggests that these galaxies do not reside in luminous galaxy clusters, it cannot be excluded that they are members of galaxy groups. To explore this possibility, we derive $3\sigma$ upper limits on the X-ray emission around these two galaxies. We computed the upper limits within a $500$~kpc region and by assuming an optically-thin thermal plasma model (\textsc{APEC} model in \textsc{XSpec}) with $kT=1.2$~keV and $Z=0.3\ \rm{Z_{\rm \odot}}$ abundance. The obtained X-ray upper limits are $<2.9\times10^{43} \ \rm{erg \ s^{-1}}$ and $<1.0\times10^{43} \ \rm{erg \ s^{-1}}$ for SDSS~J113800.88+521303.9 and SDSS~J093540.34+565323.8, respectively. These values are comparable to those obtained for galaxy groups \citep{lovisari15}, hence it cannot be excluded that these superluminous disk galaxies are members or possibly the brightest galaxies of groups.

We also probed the morphology of the five detected galaxy clusters to determine whether they are relaxed or disturbed systems. \citet{lovisari17} demonstrated that the concentration and centroid-shift are the two most robust parameters to probe the morphology of clusters. Therefore, we follow their method to measure these parameters and list the obtained values in Table \ref{tab:morphology}. We find that the galaxy cluster hosting J16273 is the most relaxed based on the concentration parameter and it is also in the range of relaxed clusters based on the centroid-shift parameter. The galaxy cluster hosting 2MASX\,J14475296+1447030 is the most relaxed cluster based on its centroid-shift. The other end of the range are represented by the clusters in the vicinity of 2MASX\,J09572689+4918571 and 2MASX\,J14072225+1352512, which exhibit disturbed morphology.

\begin{figure}[!tp]
  \begin{center}
    \leavevmode
      \epsfxsize=0.46\textwidth \epsfbox{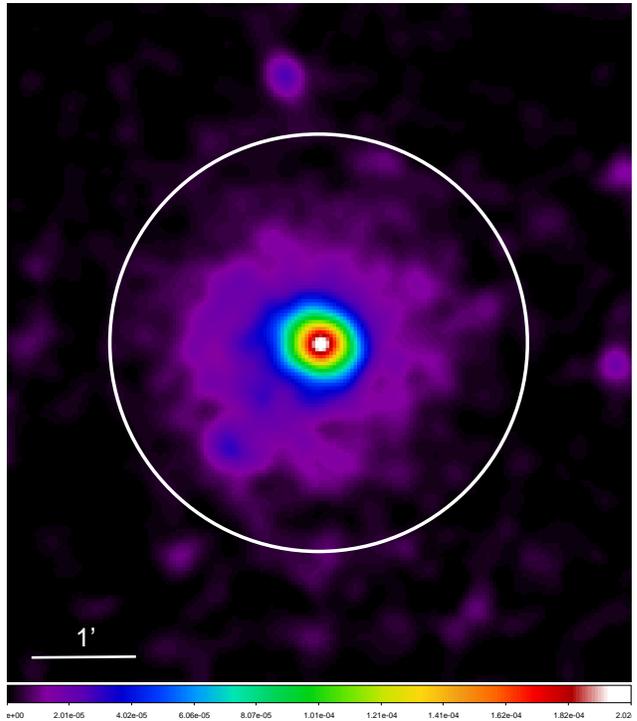}
      \vspace{0cm}
      \caption{\textit{XMM-Newton} image of the galaxy cluster in the $0.7-2$~keV band hosting J16273. The distribution of the ICM is symmetric and the cluster does not show bright surface brightness edges. The circle has a radius of $2\arcmin$ ($479$~kpc).} 
     \label{fig:clusterimg}
  \end{center}
\end{figure}

\begin{figure*}[!tp]
  \begin{center}
    \leavevmode
      \epsfxsize=0.48\textwidth \epsfbox{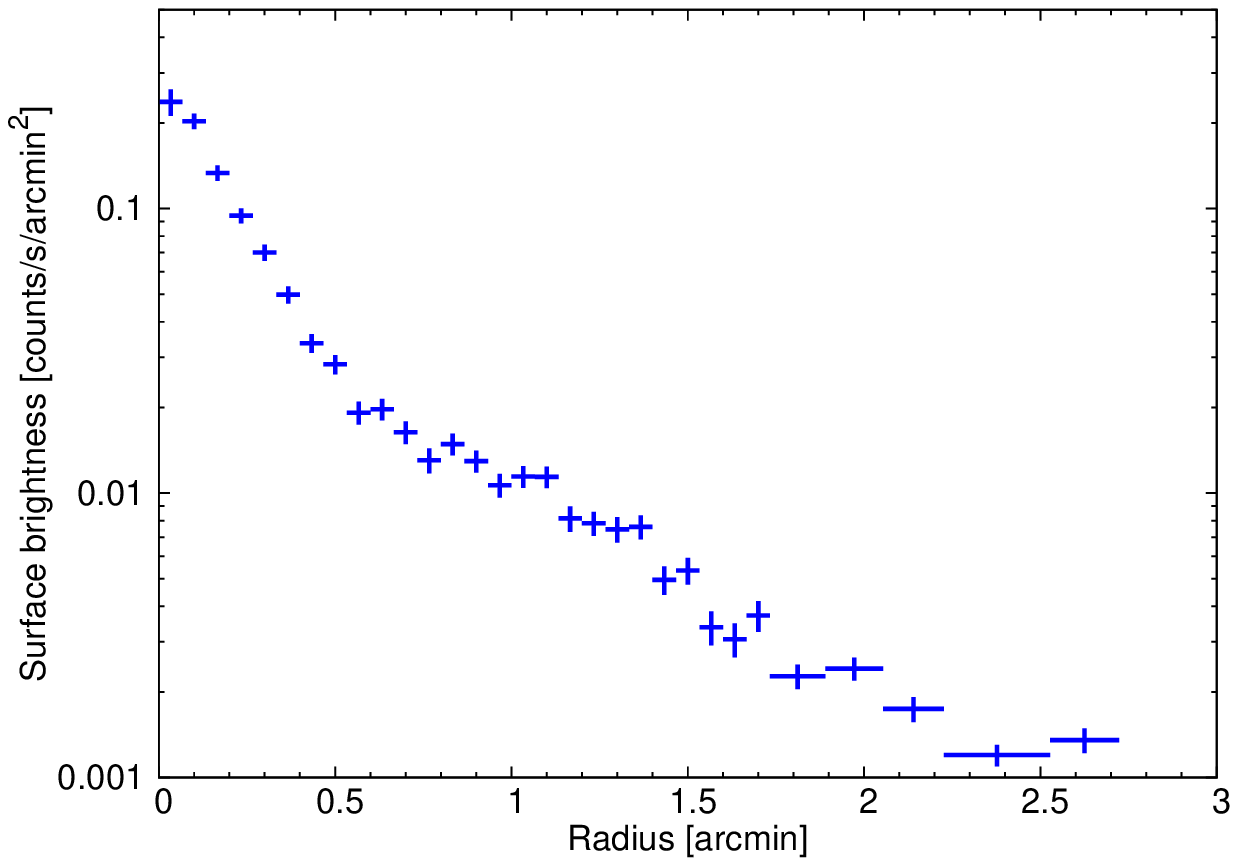}
      \epsfxsize=0.48\textwidth \epsfbox{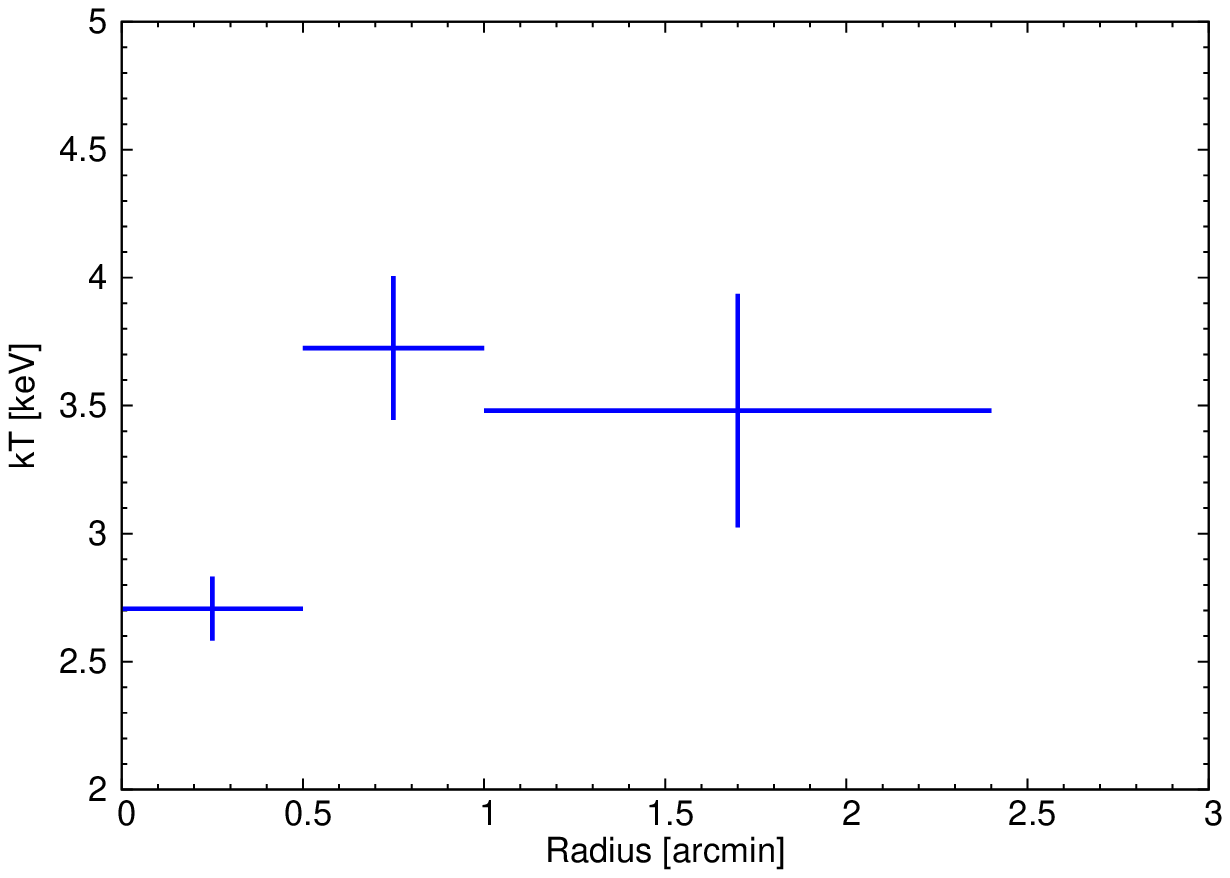}
      \vspace{0cm}
      \caption{X-ray surface brightness (left panel) and temperature (right) profile of the galaxy cluster hosting J16273. The surface brightness profile does not reveal large jumps. Note that the background is subtracted and exposure correction was applied. The temperature profile indicates lower temperature in the core, suggesting that it is a cool-core cluster.} 
     \label{fig:profiles}
  \end{center}
\end{figure*}

\begin{table}[!t]
\caption{The best-fit parameters of the gas density profile for the cluster hosting J16273}
\begin{minipage}{8cm}
\renewcommand{\arraystretch}{1.4}
\centering
\begin{tabular}{c c c c c c}
\hline 
$r_{\rm c,1}$ & $\beta_{\rm 1}$ & $n_{\rm e,1}$ & $r_{\rm c,2}$ & $\beta_{\rm 2}$ & $n_{\rm e,2}$ \\
(kpc) &  & ($\rm{cm^{-3}}$) & (kpc) &  & ($\rm{cm^{-3}}$) \\
\hline
$155.688$ & $3.058$ & $9.973\times10^{-3} $ & $196.855$ & $0.642$ & $2.473\times10^{-3}$ \\
 \hline \\
\end{tabular} 
\end{minipage}
\vspace{0.5cm}
\label{tab:density}
\end{table}

\section{Discussion}
\label{sec:discussion}

\subsection{The Galaxy Cluster with a Superluminous Spiral in its Center}
\label{sec:cluster}

While we studied the basic properties of all detected clusters in the previous section, we carried out a more in-depth analysis of the galaxy cluster that hosts a superspiral galaxy in its center. Specifically, we probe its morphology, derive surface brightness and temperature profiles, and further explore whether it is a cool-core or non-cool-core cluster.  

In Section \ref{sec:properties} we demonstrated that this galaxy cluster exhibits relaxed morphology. To further explore the morphology of the cluster, we present the $0.7-2$~keV band X-ray image in Figure \ref{fig:clusterimg}. To extract its surface brightness profile, we defined regions that have constant signal-to-noise ratios, which result in wider bins in the outskirts of the cluster.  The left panel of Figure \ref{fig:profiles} presents the $0.7-2$~keV band azimuthally averaged surface brightness profile of the cluster. The profile does not reveal sharp surface brightness edges, which often signify past interactions or mergers. This evidence suggests the relaxed nature of the cluster. 

Based on the surface brightness profile of the galaxy cluster, we also derived its density profile. To this end, we fit the surface brightness profile with a double $\beta$-model, where the surface brightness of the cluster is described following \cite{lovisari15}: 
$$ S_{\rm X} = S_{\rm 1} \Big[ 1+ \Big( \frac{r}{r_{\rm c,1}} \Big)^2 \Big]^{-3\beta_{\rm 1} +0.5} + S_{\rm 2} \Big[ 1+ \Big( \frac{r}{r_{\rm c,2}} \Big)^2 \Big]^{-3\beta_{\rm 2} +0.5} \ ,$$
where $r_{\rm c,1}$ and $r_{\rm c,2}$ are the core radii and $S_{\rm 1}$ and $S_{\rm 2}$ are the amplitudes. By fitting this model and assuming  spherical symmetry, we model the gas density profile as: 
$$ n_{\rm e}(r) = \Big \{ n_{\rm e,1}^2 \Big [ 1+ \Big ( \frac{r}{r_{\rm c,1}} \Big)^2 \Big ]^{-3\beta_{\rm 1}} + n_{\rm e,2}^2 \Big [ 1+ \Big ( \frac{r}{r_{\rm c,2}} \Big)^2 \Big ]^{-3\beta_{\rm 2}} \Big \}^{\frac{1}{2}} \ .$$
The best-fit parameters of the density profile are provided in Table \ref{tab:density}, which allows the derivation of the gas density at any radius.

In the right panel of Figure \ref{fig:profiles}, we present the temperature distribution of the cluster. To define the spectral extraction regions, we required a signal-to-noise ratio of $>30$. In the innermost region, we also required a minimum extraction region of $30\arcsec$ to reduce the impact of the point spread function of \textit{XMM-Newton}. To fit the spectrum, we used an optically-thin thermal plasma emission model (\textsc{apec} in \textsc{XSpec}), where the temperature, metallicity, and normalization were allowed to vary, but the line-of-sight column density was fixed. The profile reveals that the temperature is somewhat cooler in the core, suggesting that it is a cool-core cluster. 

Based on the iterative method introduced in \ref{sec:properties}, we found that the $R_{\rm 500}$ radius of the cluster is $797$~kpc. Within this region the best-fit temperature is $kT=3.55^{+0.18}_{-0.20}$~keV and the total ICM luminosity is $L_{\rm 500} = (1.17\pm0.12)\times10^{44} \ \rm{erg \ s^{-1}}$. Based on the $kT-M_{\rm 500}$ scaling relation, we infer a total mass of $M_{\rm 500} = (2.39 \pm 0.19) \times 10^{14} \ \rm{M_{\odot}} $. Thus, the superluminous disk galaxy, J16273, is the central BCG of a massive cluster.

\subsection{Frequency and Formation Scenarios of Superluminous Disk Galaxies as BCGs}
\label{sec:frequency}

The galaxy, J16273, is the second superluminous disk galaxy that resides in the center of a galaxy cluster. However, this galaxy exhibits a disk morphology, while 2MASX\,J10405643-0103584, presented in \citet{bogdan18b}, is a lenticular galaxy. Given that we followed up 12 candidate superluminous disk galaxy BCGs with \textit{XMM-Newton}, we can now estimate the frequency of galaxy clusters hosting such curious galaxies at their centers. 

To identify superluminous disk galaxies, \citet{ogle19} searched all galaxies in the NASA/IPAC Extragalactic Database (NED) that had available $r$-band photometry and a spectroscopic redshift of $z < 0.3$. Of these galaxies, 1525 galaxies are high-luminosity galaxies with $L_{\rm r} > 8L^{\rm \ast}$. While 1400 of these objects are elliptical galaxies, 84 and 16 galaxies exhibit spiral and lenticular morphology. The remaining 25 galaxies are either peculiar or non-spiral (e.g.\ quasar host) galaxies. As discussed in Section \ref{sec:sample} and in \citet{ogle19}, most of the superluminous disk galaxies reside in low-density environments and only a fraction of them are cluster members. Based on \citet{bogdan18b} and results of the present work, we conclude that of the 100 superluminous disk galaxies, only two systems, i.e.\ $2\%$, are the central BCGs of galaxy clusters and most cluster-member superluminous galaxies reside in the outskirts of clusters. Considering all massive galaxies with $L_{\rm r} > 8L^{\rm \ast}$ within $z<0.3$, we conclude that only $\approx0.13\%$ are superluminous disk galaxies \textit{and} central BCGs. 

The existence of massive disk galaxies is surprising because massive galaxies grow through a series of minor and major mergers, but major mergers are known to destroy galaxy disks. To explain the existence of massive disk galaxies in the present universe, two formation channels are considered. First, it is possible that a fraction of superluminous disk galaxies experienced a very quiet merger history. \citet{ogle19} analyzed the environment of the superluminous disk galaxies and concluded that most of them reside in relatively low-density environments. In agreement with this, our study also pointed out that only $\sim2\%$ of superluminous disks reside in the center of rich clusters. Therefore, it is feasible that a substantial fraction of superluminous disk galaxies did not undergo major mergers and gradually increased their mass by a series of minor mergers. Additionally, due to their large mass, these galaxies are protected from major mergers since most galaxies in their (low-density) vicinity will have significantly lower masses, resulting in minor mergers. 

In the second formation scenario, the original galaxy disks are destroyed, but they are re-formed at a later epoch. Because of the high galaxy density in the central regions of clusters, the occurrence rate of major mergers is significantly higher, which in turn is more likely to result in the formation of a spheroid. Therefore, in these environments, it is more likely that the galaxy disks were accreted at a later time \citep{jackson20}. In this picture, the presence of an extended disk can be explained by the merger between a spheroid and a gas-rich satellite, where the latter galaxy spins up the spheroid. As a result of this merger, a rotationally supported disk can form. According to simulations presented in \citet{jackson20}, about $70\%$ of massive disk galaxies originate from this scenario, and only $30\%$ of them can be attributed to the quiet merger history. 

Given the rich environment of cluster cores, it is possible that J16273 also has a re-formed disk. This scenario is also consistent with the relatively massive bulge component of the galaxy, which bulge would be sufficiently massive as traditional spheroid even without its large disk. In \citet{bogdan18b} we also speculated that the massive superluminous lenticular galaxy with $B/T=0.66$, 2MASX\,J10405643-0103584, was an elliptical galaxy whose disk was re-formed by a gas-rich merger. Because both superluminous disk galaxies in the cluster cores have relatively large $B/T$ ratios, we hypothesize that re-forming the disk is a more likely scenario to result in the formation of these galaxies. However, in isolated environments or in cluster outskirts, massive disk galaxies could have formed via a quiet merger history.

\subsection{Outlook}
\label{sec:outlook}

In the present study, we probed whether superluminous disk galaxies are BCGs of clusters. However, our analysis could not be extended to galaxy groups. Indeed, our initial selection criteria utilized short \textit{ROSAT} observations to detect the ICM emission from clusters. Although these data are suitable to hint the presence of luminous galaxy clusters at $z\sim0.2-0.3$, fainter galaxy groups remain hidden at these redshifts. Therefore, it is feasible that some of the superluminous disk galaxies are located in galaxy groups or in low-mass clusters. 

To probe superluminous disk galaxies residing in such systems, the \textit{eROSITA} All-Sky Survey will be ideal. Thanks to its large collective area, after the 4-year survey it will detect galaxy groups with a few times $10^{42} \ \rm{erg \ s^{-1}}$ luminosity within $z<0.3$. Although its point spread function is significantly broader than that of \textit{XMM-Newton}, these data could still be used to identify the centers of groups. Thus, these data will allow us to probe whether superluminous disks are members of galaxy groups or low-mass clusters, and will reveal whether they reside in the core or outskirts of these systems. 

\section{Summary}
\label{sec:summary}

In this work, we explored the environment of a sample of superluminous disk galaxies using \textit{XMM-Newton} observations. Our results can be summarized as follows: 
\begin{itemize}
    \item We probed the environment of seven superluminous disk galaxies. We detected luminous galaxy clusters with $kT = 2.5-4.7$~keV in the vicinity of five systems. 
    \item We established that one superluminous disk galaxy, J16273, resides in the center of a galaxy cluster. The other superluminous disk galaxies either reside in the outskirts of clusters or are not physically associated with any clusters.
    \item We characterized the galaxy cluster that hosts a superluminous spiral in its center and established that it has $kT=3.55^{+0.18}_{-0.20}$~keV and $M_{\rm 500} = (2.39 \pm 0.19) \times 10^{14} \ \rm{M_{\odot}} $. 
    \item We identify the central galaxies of the four clusters, in which the superluminous galaxies have large offsets. The central galaxies of these clusters are massive ellipticals. However, for two clusters the superluminous disk galaxy in the outskirt is more luminous than the central galaxy, making them  offset brightest cluster galaxies.
    \item We established that $\sim2\%$ of superluminous disk galaxies reside in cluster centers and only $\approx0.13\%$ of massive ($L_{\rm r} > 8L^{\rm \ast}$) galaxies retained their disk morphology in the center of a cluster. This is likely because frequent major mergers destroy galaxy disks. We speculate that the disk of superluminous spiral and lenticular galaxies in the center of clusters may have been re-formed due to a merger with a gas-rich satellite. 
\end{itemize}

\smallskip

\begin{small}
\noindent
\textit{Acknowledgements.}
We thank the referee for his/her constructive report. This work uses observations obtained with \textit{XMM-Newton}, an ESA science mission with instruments and contributions directly funded by ESA Member States and NASA. In this work, the NASA/IPAC Extragalactic Database (NED) have been used. Funding for SDSS-III has been provided by the Alfred P. Sloan Foundation, the Participating Institutions, the National Science Foundation, and the U.S. Department of Energy Office of Science. The SDSS-III web site is http://www.sdss3.org/. SDSS-III is managed by the Astrophysical Research Consortium for the Participating Institutions of the SDSS-III Collaboration including the University of Arizona, the Brazilian Participation Group, Brookhaven National Laboratory, Carnegie Mellon University, University of Florida, the French Participation Group, the German Participation Group, Harvard University, the Instituto de Astrofisica de Canarias, the Michigan State/Notre Dame/JINA Participation Group, Johns Hopkins University, Lawrence Berkeley National Laboratory, Max Planck Institute for Astrophysics, Max Planck Institute for Extraterrestrial Physics, New Mexico State University, New York University, Ohio State University, Pennsylvania State University, University of Portsmouth, Princeton University, the Spanish Participation Group, University of Tokyo, University of Utah, Vanderbilt University, University of Virginia, University of Washington, and Yale University. \'A.B., C.J., and W.R.F acknowledge support from the Smithsonian Institution and the Chandra High Resolution Camera Project through NASA contract NAS8-03060. O.E.K is supported by the GA\v{C}R EXPRO grant No. 21-13491X.  L.L. acknowledges financial contribution from the contracts ASI-INAF Athena 2019-27-HH.0, ``Attività di Studio per la comunità scientifica di Astrofisica delle Alte Energie e Fisica Astroparticellare'' (Accordo Attuativo ASI-INAF n. 2017-14-H.0), and from INAF ``Call per interventi aggiuntivi a sostegno della ricerca di main stream di INAF''. 
\end{small}

\bibliographystyle{apj}
\bibliography{paper2.bib} 

\end{document}